\documentclass[pra,superscriptaddress,column]{revtex4}
\usepackage[a4paper, total={5.7in, 8in}]{geometry}
\usepackage{optidef}

\usepackage{graphicx,bm,natbib,upgreek,amsmath,mathrsfs,accents}
\usepackage{amsbsy}
\usepackage{amsfonts}

\usepackage{amsthm}

\usepackage[dvipsnames]{xcolor}

\usepackage{graphicx}
\usepackage{subfigure}

\begin{document}

\title{Spatial-mode-demultiplexing for enhanced intensity and distance measurement}

\author{Luigi Santamaria \footnote{luigi.santamaria@asi.it}}
\author{Deborah Pallotti}
\author{Mario Siciliani de Cumis}
\author{Daniele Dequal}
\affiliation{Agenzia Spaziale Italiana, Matera Space Center, Contrada Terlecchia snc. 75100 Matera, Italy}

\author{Cosmo Lupo \footnote{cosmo.lupo@poliba.it}}
\affiliation{Dipartimento di Fisica, Politecnico di Bari \& Universit\`a di Bari, 70126, Bari, Italy}
\affiliation{INFN, Sezione di Bari, 70126 Bari, Italy}

\begin{abstract}
Spatial-mode demultiplexing (SPADE) has recently been adopted to measure the separation in the transverse plane between two incoherent point-like sources. It has been argued that this approach may yield extraordinary performances in the photon-counting regime. Here, we explore SPADE as a tool for precision measurements in the regime of bright, incoherent sources. First we analyse the general problem of estimating the second moments of the source's intensity distribution, for an extended incoherent source of any shape. Second, we present an experimental application of SPADE to the case of two point-like, bright sources. We demonstrate the use of this setup for the estimation of the transverse separation and for the estimation of their relative intensity. 
\end{abstract}

\maketitle


\section{Introduction}

Over the time several criteria have been developed to define, in a more or less quantitative way, the resolving power of an optical imaging system. 
The Rayleigh criterion is among the most popular ones \cite{Rayleigh1879}. It states, heuristically, that the wavelength sets the minimum resolvable transverse separation between two point-like sources.
Consider an optical system characterised by its point-spread function (PSF), whose width is of the order of the Rayleigh length $\mathrm{x_R} = \lambda D/R$, where $\lambda$ is the wavelength, $D$ is the distance to the object, and $R$ is size of the pupil.
According to the Rayleigh criterion, two quasi-monochromatic point-sources are hard to resolve when their transverse separation $s$ is comparable to or smaller than $\mathrm{x_R}$.
This criterion is strictly related to the PSF of the optical system, however, in practice the knowledge of the PSF is not sufficient to determine the resolution power, which also depends on signal-to-noise ratio (SNR).
A number of techniques have been developed to beat the Rayleigh limit \cite{Hell2007}. These includes switching the emission on and off \cite{Dickson1997}, near field probing \cite{Drig1986}, or exploiting optical non-linearities \cite{Hell1994}, just to name a few.
Most of these techniques rely on source engineering, which is not an option for astronomical observations.

Recently, Tsang at al.~\cite{Tsang2016} analysed this classical resolution limit using tools from quantum estimation theory. 
They formulated the problem of estimating the transverse separation between two incoherent point-like sources, and showed that the separation can be estimated with constant precision even when $s \ll \mathrm{x_R}$, which implies far-field sub-Rayleigh resolution.
Quantum estimation theory applies to the photon-counting regime, where the probability of obtaining more than one photon per detection event is highly suppressed. 
This result is enabled by linear optics and photon detection, and exploits the additional information contained in the phase and in the spatial correlations of the optical field.
Such information is ignored in direct imaging (DI) but can be extracted through a structured measurement that allows for coherent processing of the field before detection. Examples are interferometric techniques such as SPAtial mode DE-multiplexing (SPADE) \cite{Zhou2019,Xue2001,Abouraddy2012,Martin2017} or Super-Localization via Image-inVERsion interferometry (SLIVER)~\cite{Larson2019,Tang2016,Wicker2009}.
These and other interferometric techniques \cite{Tham2017,PRL2020,Ugo2022} have been demonstrated for super-resolution imaging and high-precision distance measurements \cite{Parniak2018, Sorelli}, especially for the problem of estimating the transverse separation between two point-sources \cite{Par2016,Tham2017,Yang2016,Boucher2020}, both in the photon-counting regime~\cite{MT2019} and for bright sources~\cite{Lvovsky,Nayak}.

These results eventually suggested a modern re-formulation of the Rayleigh criterion~\cite{Modern}, which applies to 1D and 2D incoherent sources of any shape, not necessarily a pair of point-sources. %
It is possible to see that a structured measurement allows us to estimate with constant precision the first and second moments of the source's intensity distribution (i.e.~independently of the value of the ratio $\gamma:= s/\mathrm{x_R}$, where $s$ is the size of the source)~\cite{MankeiPRA,Modern} --- matching the ultimate quantum bound.
On the other hand, the precision in the estimation of higher moments goes to zero with $\gamma$, yet with better scaling than DI.
Analogous results were obtained by Grace and Guha for the problem of discriminating extended sources with different shapes~\cite{GraceGuha}. These authors showed that, compared with DI, SPADE improves the scaling of the error exponent with the parameter $\gamma$, up to saturation of the ultimate quantum limit.
Similar conclusions were drawn in Refs.~\cite{Shapiro,PRL2021} for the case of two point-sources. In particular, Ref.~\cite{PRL2021} considered the problem of discriminating the presence of a weak secondary sources in the vicinity of a brighter one, suggesting application to exoplanet detection and spectroscopy \cite{astro22}. 

Inspired by the results obtained in the photon-counting regime, here we explore the use of a structured measurement, in particular SPADE, for realising precision measurements in the regime of bright, incoherent sources.
The latter is fundamentally different from that of weak sources.
In our experimental setup, we observe fast random fluctuations due to mechanical vibrations and slower thermal fluctuations.
We perform repeated measurements by changing the separation between the sources, which leads to hysteresis effects. This induces statistical errors due to the fact that movable mechanical parts do not come back exactly to the same position. This latter repeatability error dominates the measurement uncertainty in our experimental setup.
By contrast, measurements in the regime of weak signals are dominated, at least in principle, by shot noise \cite{Tsang2016}. 
In practice, in both regimes the detection limit of SPADE is determined by cross talk~\cite{PhysRevLett.125.100501,Boucher2020,Linowski}.


In this work we consider a model of extended incoherent bright source (Section \ref{Sec:model}).
First we present a general error analysis for the estimation of the second moments for sources of any shape (Section \ref{Sec:error}). We argue that in general SPADE allows for improved precision compared to DI.
Second, we present an experimental setup for the case of two point-like sources of different brightness
(Section \ref{Sec:exp}).
We demonstrate the use of SPADE to achieve precision measurement of the transverse source separation (this was also discussed by Boucher \textit{et al.}~\cite{Boucher2020} for the case of equal brightness, with a particular focus on the error due to cross-talk)
and of the relative intensity.


\section{The model}\label{Sec:model}

Consider an extended two-dimensional source of incoherent, quasi-monochromatic light of wavelength $\lambda$. 
The source lays in the object plane, and an optical system is used to create a focused image in the image plane.
The optical system is characterised by its PSF, which describes the field in the image plane generated by a point-source located in the object plane.
In a scalar theory, within the far-field and paraxial approximation, a point-like emitter at position $\mathbf{r}_o=(u,v)$ in the object plane yields an optical field in the image plane proportional to the PSF $T$, where~\cite{goodman2008introduction}
\begin{align}\label{Fourier}
T(\mathbf{r}_i-M \mathbf{r}_o) = \mathcal{N}
\iint d^2 \mathbf{r} \, P(\mathbf{r}) e^{-i 2\pi \frac{\mathbf{r}_i - M \mathbf{r}_o}{\lambda M d} \cdot \mathbf{r}} \, .
\end{align}
Here $\mathbf{r}_i=(x,y)$ is the coordinates vector in the image plane, $M$ is the magnification factor, and $P$ is the pupil function depending on Cartesian coordinates $\mathbf{r}=(t,z)$.
We choose the factor $\mathcal{N}$ in such a way that the PSF is normalised to one.
Without loss of generality the PSF may be assumed to be real-valued~\cite{Modern,PhysRevLett.117.190802,Tsang2016,goodman2008introduction}.

An extended incoherent source is characterised by a distribution of mutually-incoherent point-like sources in the object plane, with position-dependent intensity $I_{uv}$.
DI consists in a direct measurement of the intensity $I(x,y)$ on each point $(x,y)$ of the image plane. Since the source is incoherent we have
\begin{align}
    I(x,y) = \sum_{u,v} I_{uv} \, |T(\mathbf{r}_i-M \mathbf{r}_o)|^2 \, .
\end{align}

The aim of this Section is to compare DI with a structured detection strategy where the field in the image plane is first demultiplexed in its transverse degrees of freedom, and then measured.
Such method of SPADE may be implemented in a number of ways, also depending on the basis used for spatial demultiplexing in the image plane.

Given the PSF, consider its derivatives 
\begin{align}
\frac{\partial^{a+b} T}{\partial x^a \partial y^b}(\mathbf{r}_i) \, ,
\end{align}
for $a,b \geq 0$, and use them to define an orthonormal basis by applying the Gram-Schmidt orthogonalisation procedure.
The procedure can be simplified if the PSF has symmetry.
For example, if the PSF is even along each Cartesian axis, i.e., such that $T(x,y)=T(-x,y)=T(x,-y)$, the functions 
$\frac{\partial^{a+b} T}{\partial x^a \partial y^b}$ and $\frac{\partial^{c+d} T}{\partial x^c \partial y^d}$ are mutually orthogonal unless $a$ has the same parity as $c$, and $b$ has the same parity as $d$. 
\footnote{In general, the PSF inherits the symmetry of the pupil function in Eq.~(\ref{Fourier}). For examples, an even PSF is obtained from  a rectangular pupil, or for a circular one, which yield PSFs expressed by Bessel functions, whereas a Gaussian PSF is obtained from a Gaussian pupil function.}
For a PSF with this symmetry, the Gram-Schmidt orthogonalisation procedure yields
\begin{align}
    f_{00}(\mathbf{r}_i) & := T(\mathbf{r}_i) \, , \label{f00} \\
    f_{01}(\mathbf{r}_i) & := \mathcal{N}_{01} \frac{\partial T}{\partial x}(\mathbf{r}_i) \, , \label{f01} \\
    f_{10}(\mathbf{r}_i) & := \mathcal{N}_{10} \frac{\partial T}{\partial y}(\mathbf{r}_i) \, , \label{f10} \\
    f_{11}(\mathbf{r}_i) & = \mathcal{N}_{11} \frac{\partial^2 T}{\partial x \partial y}(\mathbf{r}_i) \, , \label{f11} 
\end{align}
and so on and so forth.
The factors $\mathcal{N}_{hk}$ are chosen in such a way that the functions $f_{hk}$ are normalised to one.
Here we focus only on the lowest order terms (\ref{f00})-(\ref{f10}).

SPADE can be implemented experimentally in a number of ways, e.g.,~with an interferometer ~\cite{MT2019}, a hologram~\cite{Par2016}, a multimode wave guide~\cite{Tsang2016}.
The measurement output is the intensity of the field along each function (mode) in the basis:
\begin{align}
    I(f_{hk}) = \sum_{u,v} I_{uv} 
    \left| \langle f_{hk},T_{uv} \rangle \right|^2 \, , 
\end{align}
where $\langle f_{hk},T_{uv} \rangle$ is a short-hand notation for 
$\iint dx dy \, f_{hk}^*(\mathbf{r}_i) T(\mathbf{r}_i-M\mathbf{r}_0)$, and ${}^*$ indicates complex conjugation.

We are especially interested in the sub-diffraction regime where the object size (multiplied by the magnification factor $M$) is much smaller than the width of the PSF, where the latter is quantified by the Rayleigh length $\mathrm{x_R}$.
In this regime we can expand the PSF up to the first order in $u$ and $v$,
\begin{align}
    T(\mathbf{r}_i-M\mathbf{r}_0) & \simeq T(\mathbf{r}_i) + M u \frac{\partial T}{\partial x}(\mathbf{r}_i)
    + M v \frac{\partial T}{\partial y}(\mathbf{r}_i) \, ,
\end{align}
which in turn implies 
\begin{align}
    I(f_{01}) & = \sum_{u,v} I_{uv} 
    \left| \langle f_{01} , T_{uv} \rangle \right|^2 \\
    & \simeq \sum_{u,v} I_{uv} 
    \left| \langle f_{01} , f_{00} \rangle 
    + \frac{M u}{\mathcal{N}_{01}} \langle f_{01} , f_{01} \rangle
    + \frac{M v}{\mathcal{N}_{10}} \langle f_{01} , f_{10} \rangle
    \right|^2 \\
    & = \sum_{u,v} I_{uv} 
    \left| \frac{M u}{\mathcal{N}_{01}} \langle f_{01} , f_{01} \rangle
    \right|^2 
    = \frac{M^2}{\mathcal{N}_{01}^2} \sum_{u,v} I_{uv} u^2 \, ,
\end{align}
and similarly
\begin{align}
    I(f_{10}) \simeq \frac{M^2}{\mathcal{N}_{10}^2} \sum_{u,v} I_{uv} v^2\, .
\end{align}
This shows that measuring the intensity in the first order modes $f_{01}$ and $f_{10}$ yields, in the sub-diffraction regime, a direct measurement of the second order moments of the source intensity distribution.
This result can be extended to higher moments, for example from
\begin{align}
    f_{11} = \mathcal{N}_{11} \frac{\partial^2 T}{\partial x \partial y}
\end{align}
we obtain
\begin{align}
    I(f_{11}) \simeq \sum_{u,v} I_{uv} \left| 
    M^2 u v \langle f_{11} , \frac{\partial^2 T}{\partial x \partial y} \rangle
    \right|^2
    = \frac{M^4}{\mathcal{N}_{11}^2}  \sum_{u,v} I_{uv} u^2 v^2
    \left| 
    \langle f_{11} , f_{11} \rangle
    \right|^2
    = \frac{M^4}{\mathcal{N}_{11}^2} \sum_{u,v} I_{uv} u^2 v^2
    \, .
\end{align}

Expanding Eq.~(\ref{Fourier}) up to the second order in $(\mathbf{r}_i - M \mathbf{r}_o) \cdot \mathbf{r} /(\lambda M d)$ we obtain
\begin{align}
T(\mathbf{r}_i-M \mathbf{r}_o) \simeq \mathcal{N}
\iint d^2 \mathbf{r} \, P(\mathbf{r}) \left[
1 - i 2\pi \frac{\mathbf{r}_i - M \mathbf{r}_o}{\lambda M d} \cdot \mathbf{r}
- 4\pi^2 \left( \frac{(\mathbf{r}_i - M \mathbf{r}_o) \cdot \mathbf{r}}{\lambda M d } \right)^2
\right] \, .
\end{align}
If the pupil function is even along each Cartesian axis, then the first-order term vanishes,
\begin{align}
T(\mathbf{r}_i-M \mathbf{r}_o) \simeq \mathcal{N}
\iint d^2 \mathbf{r} \, P(\mathbf{r}) \left[
1 - 4\pi^2 \left( \frac{(\mathbf{r}_i - M \mathbf{r}_o) \cdot \mathbf{r}}{ \lambda M d } \right)^2
\right] \, .
\end{align}
Putting $\iint d^2 \mathbf{r} \, P(\mathbf{r}) = 1$ and if the pupil functions is invariant under the interchange of the Cartesian axes, $P(w,z) = P(z,w)$, this further simplifies to
\begin{align}
T(\mathbf{r}_i-M \mathbf{r}_o) 
& \simeq \mathcal{N}
\left[
1 - 4\pi^2 \frac{(x -M u)^2 \tau^2 + (y -M )^2 \tau^2 }{ (\lambda M d)^2 }
\right] \\
& \simeq \mathcal{N} e^{- 4\pi^2 \frac{(x -M u)^2 \tau^2 + (y -M v)^2 \tau^2 }{ (\lambda M d)^2 }}
\, ,
\end{align}
where $\tau^2$ is the second moment of the pupil function, $\tau^2 = \iint d^2\mathbf{r} P(\mathbf{r}) t^2
= \iint d^2\mathbf{r} P(\mathbf{r}) z^2$.
This shows that, if the PSF has enough symmetry, it can be approximated by a Gaussian function in the sub-Rayleigh regime.
Finally, note that $\tau$ is of the order of the size the pupil $R$. Therefore, the standard deviation $\lambda D/\tau$ plays the role of the Rayleigh length.


\section{Error analysis}\label{Sec:error}

In this Section we focus on the problem of estimating the second moments of the source's intensity distribution.
To compare DI with SPADE, we need to estimate the measurement uncertainty in each case.
To make things more concrete, we consider in detail the case of a Gaussian PSF.
Therefore, the point-source located at position $(u,v)$ in the object plane yields, on the image plane, an electric field with Gaussian envelope
(to simplify the notation we put $M=1$)
\begin{align}
    S_{uv}(x,y) & = \sqrt{\frac{I_{uv}}{2\pi w^2}} \, e^{-\frac{(x-u)^2+(y-v)^2}{4 w^2}} \, ,
\end{align}
where the standard deviation plays the role of the Rayleigh length, i.e., 
$w \equiv \mathrm{x_R}$.
For a Gaussian PSF, the basis in Eq.~(\ref{f00})-(\ref{f11}) is given by the Hermite-Gaussian (HG) functions, $f_{hk} \equiv \text{HG}_{hk}$.

Since the point-sources are incoherent, measuring the intensity of the field component in the lowest order modes yields
\begin{align}
    I(\text{HG}_{01}) 
    & =
    \sum_{u,v}
    I_{uv} \, 
    \frac{u^2}{4 w^2} \, e^{ - \frac{u^2+v^2}{4 w^2}} 
    \simeq 
    \frac{1}{4 w^2} \sum_{u,v} I_{uv} u^2
\, , \\
    I(\text{HG}_{10}) 
    & =
    \sum_{u,v} 
    I_{uv} \, \frac{v^2}{4 w^2} \, e^{ - \frac{u^2+v^2}{4 w^2}}
    \simeq 
        \frac{1}{4 w^2} \sum_{u,v} I_{uv} v^2
    \, .
\end{align}
As discussed above, in the sub-Rayleigh regime this allows us to estimate the second moments of the source intensity distribution.

We denote the second moments as $\mu_{u^2}$ and $\mu_{v^2}$. We then have
\begin{align}
    \mu_{u^2} & \simeq 4 w^2 I(\text{HG}_{01}) \, , \\
    \mu_{v^2} & \simeq 4 w^2 I(\text{HG}_{10}) \, .
\end{align}
Assuming that $w$ is know with negligible error (i.e.,~the optical system is well characterised) and assuming for now ideal SPADE with no cross talk, the error in the estimation of the moments is dominated by the error in the measurement of the intensity 
$I(\text{HG}_{01})$, $I(\text{HG}_{10})$.
We expect this error to be proportional to the square root of the intensity, therefore the SNR is (for example for the measurement of $\mu_{u^2}$):
\begin{align}\label{SNR_SPADE}
\text{SNR}^{(\text{SPADE})} \sim \sqrt{I(\text{HG}_{01})} \, . 
\end{align}

What happens if we estimate the second moments by DI?
We would measure the intensity pixel by pixel:
\begin{align}\label{int_DI}
    I(x,y) = \sum_{u,v} |S_{uv}(x,y)|^2 
    = \sum_{u,v} 
    \frac{I_{uv}}{2\pi w^2}
    e^{-\frac{(x-u)^2+(y-v)^2}{2 w^2}} \, ,
\end{align}
and then use this data to compute the moment of the field on the image screen. For example the second moment in the $x$ direction is
\begin{align}
    \mu_{x^2} 
    & = \iint dx dy \, x^2 \sum_{u,v} |S_{uv}(x,y)|^2 \\ 
    & = \iint dxdy \, x^2 \sum_{u,v} \frac{I_{uv}}{2\pi w^2}
    e^{-\frac{(x-u)^2+(y-v)^2}{2 w^2}} \\
    & = \sum_{u,v} I_{uv}
    \iint dxdy \, x^2  \frac{1}{2\pi w^2}
    e^{-\frac{(x-u)^2+(y-v)^2}{2 w^2}} \\
        & = \sum_{u,v} I_{uv}
    \int dx \, x^2  \sqrt{\frac{1}{2\pi w^2}}
    e^{-\frac{(x-u)^2}{2 w^2}} \\
    & = \sum_{u,v} I_{uv}
    ( w^2 + u^2 )
    =
    I w^2 + \mu_{u^2} \, ,
\end{align}
where $I = \sum_{uv} I_{uv}$ is the total intensity.
From this we can finally obtain the moment of the source by subtracting the term $I w^2$.

In the sub-Rayleigh regime, Eq.~(\ref{int_DI}) yields 
$I(x,y) \simeq I/ (2\pi w^2)$.
The error in the measurement of the intensity is expected to be proportional to $\sqrt{I(x,y)}$, 
whereas the quantity to be estimated, the second moment, is proportional to $I(\text{HG}_{01})$.
Therefore the SNR for DI is
\begin{align}\label{SNR_DI}
\text{SNR}^{(\text{DI})}
\sim \frac{I(\text{HG}_{01})}{\sqrt{I}} \, ,
\end{align}
This is a factor $\sqrt{I(\text{HG}_{01})/I}$ smaller than Eq.~(\ref{SNR_SPADE}), which is much smaller than one in the sub-Rayleigh regime.

In conclusion, the relation between the SNR's in Eqs.~(\ref{SNR_SPADE}) and~(\ref{SNR_DI}) expresses the quantitative advantage of SPADE over DI. Intuitively, this advantage comes from the fact that SPADE allows us to estimate the second moment with a single intensity measurement, whereas DI gives local information on the intensity in each pixel in the image plane --- which we then need to post-process to obtain the moment.


\subsection{Two point-like sources}

Consider in more detail the case of two point-like sources. The primary source (A) has intensity $I_A$, and the secondary source (B) has intensity $I_B < I_A$, and $d$ is their transverse separation in the object plane.
The optical system is aligned with the primary source and has a Gaussian PSF with width $w$.
Therefore, the field intensity on the image plane is
\begin{align}
I(x,y) = \frac{I_A}{2\pi w^2}\, e^{-\frac{x^2+y^2}{2 w^2}}
+ \frac{I_B}{2\pi w^2}\, e^{-\frac{(x-d \cos{\theta})^2+(y-d\sin{\theta})^2}{2 w^2}} \, ,
\end{align}
which can be measured by DI, where $\theta$ is the angle of the secondary sources with respect to the reference frame.

The lowest order modes in the HG basis are 
\begin{align}
    \text{HG}_{00}(\mathbf{r}_i) & := \sqrt{\frac{1}{2\pi w^2} } \, e^{-\frac{x^2+y^2}{4 w^2}} \, , \label{HG00} \\
    \text{HG}_{01}(\mathbf{r}_i) & := \sqrt{\frac{1}{2\pi w^2} } \, \frac{x}{w} \, e^{-\frac{x^2+y^2}{4 w^2}} \, , \label{HG01} \\
    \text{HG}_{10}(\mathbf{r}_i) & := \sqrt{\frac{1}{2\pi w^2} } \, \frac{y}{w} \, e^{-\frac{x^2+y^2}{4 w^2}} \, . \label{HG10} 
\end{align}
A SPADE measurement in this basis, in the sub-Rayleigh regime, yields
\begin{align}
    I(\text{HG}_{00}) & := I_A 
    + I_B \, e^{- \frac{d^2}{4 w^2}} 
    \simeq I_A + I_B 
    \, , \label{I_HG00} \\
    I(\text{HG}_{01}) & := I_B \, \frac{d^2 (\cos{\theta})^2}{4 w^2} \, e^{- \frac{d^2}{4 w^2}} 
    \simeq 
    I_B \, \frac{d^2 (\cos{\theta})^2}{4 w^2} 
    \, , \label{I_HG01} \\
    I(\text{HG}_{10}) & := I_B \, \frac{d^2 (\sin{\theta})^2}{4 w^2} \, e^{- \frac{d^2}{4  w^2}} 
    \simeq 
    I_B \, \frac{d^2 (\sin{\theta})^2}{4 w^2} 
    \, . \label{I_HG10} 
\end{align}
From a measurement of $I(\text{HG}_{01})$ and $I(\text{HG}_{10})$ we can estimate up to two of the three parameters, $I_B$, $d$, and $\theta$, that characterise the secondary source.
In principle an ideal SPADE measurement would allow us to completely decoupled $I_B$ from $I_A$.
In practice, this decoupling is only partial due to cross talk.


\subsection{Cross talk}

In practice, experimental implementations of SPADE are subject to cross talk between the HG channels. 
This is due to inevitable imperfections in the manufacture of the device.
The cross talk between $\text{HG}_{00}$ and $\text{HG}_{nm}$ is quantified by the parameter
\begin{equation}
    x_{nm}^{00}=10 \log_{10}\left(\frac{P_{nm}}{P_{00}}\right)
    \, ,
\end{equation}
where $P_{nm}$ represents the  power on the $\text{HG}_{nm}$ output channel when only $\text{HG}_{00}$ is injected with a power equal to $P_{00}$ from the  free space input (in our device the input is in free space and all output channels are fiber coupled).
The limitation is due to the presence of radiation in high order modes even when the incoming light is fully matched with the demultiplexer (i.e.,~when in principle only $\text{HG}_{00}$ should be excited). 
Cross talk between generic modes 
$\text{HG}_{n'm'}$ and $\text{HG}_{nm}$
also exists, but in the sub-Rayleigh regime the strongest signal comes from the fundamental mode $\text{HG}_{00}$.
The cross talk induces both systematic and stochastic errors. The offset values, reported in Table \ref{ct1}, represent systematic errors that in principle can be compensated for. 
These values are in turn subject to random fluctuations due to intrinsic nature of the light as well as mechanical vibrations, thermal fluctuations, and hysteresis effects in the measurement apparatus. 

\begin{table}[t!] 
\center
\begin{tabular}{|p{1cm}||p{1cm}|p{1cm}|p{1cm}|p{1cm}|p{1cm}|   }
 \hline
 dB & $x_{01}^{00}$ & $x_{10}^{00}$  & $x_{02}^{00}$  & $x_{20}^{00}$ & $x_{11}^{00}$  \\
 \hline
Spec & -26   & -30    & -29 & -30 & -28\\
  \hline
  
Exp & -26 $\pm 1$   & -30 $\pm 1$   & -28 $\pm 1$& -29 $\pm 1$ & -29 $\pm 1$\\
  \hline
 \end{tabular}
\caption{Cross-talk values from $\text{HG}_{00}$ mode into the $\text{HG}_{nm}$ modes. 
We compare the values given in the product specification with the measured values.
{The cross talk values were measured by injecting a $\text{HG}_{00}$ beam with a waist fully matched with demultiplexer ($w_0=300 \mu m$) in free-space input, recording the power $P_{nm}$  in all $\text{HG}_{nm}$ output channels and calculating $x_{nm}^{00}=10 \log_{10}\left(\frac{P_{nm}}{P_{00}}\right)$. We considered an uncertainty of 1 dB given by the discrepancy between measured cross talk (Exp) and cross talk values from datasheet (Spec).}} \label{ct1}
\end{table}

As a matter of fact, cross talk limit the SNR.
Consider the estimation of $\mu_{u^2}$ via a measurement of $I(\text{HG}_{01})$. 
The noise is due to fluctuations not only in the signal, but also in the $\text{HG}_{00}$ component due to cross talk.
Taking this into account, the SNR reads
\begin{align}
   \text{SNR}^{(\text{SPADE})} 
   \sim \frac{I(\text{HG}_{01})}{\sqrt{I(\text{HG}_{00})10^{x_{01}^{00}/10} +I(\text{HG}_{01})}} 
   \sim \frac{I_B}{\sqrt{ I_A 10^{x_{01}^{00}/10} + I_B }  } \, .
\end{align} 
Note that the SNR goes as $\sqrt{I(\text{HG}_{01})} \sim \sqrt{I_B}$ only if the cross talk is negligible, in which case we recover Eq.~(\ref{SNR_SPADE}).

Let us compare this with DI.
In the sub-Rayleigh regime, the two sources largely overlap, yielding
\begin{equation}
   \text{SNR}^{(\text{DI})} = \frac{I_B}{\sqrt{I_A + I_B} }  \, ,
\end{equation} 
which is analogous to Eq.~(\ref{SNR_DI}).
For example, in our experimental setup we have,
$I_A=33.6 \mu W$ and $I_B=0.9 n W$, 
which yield
\begin{equation}     
\frac{\text{SNR}^{(\text{SPADE})}}{\text{SNR}^{(\text{DI})}} \simeq 20 \, .
\end{equation} 
That is, we expect an SNR improvement of one order of magnitude for SPADE over DI. 


\begin{figure}
    \centering
    \includegraphics[scale=.35]{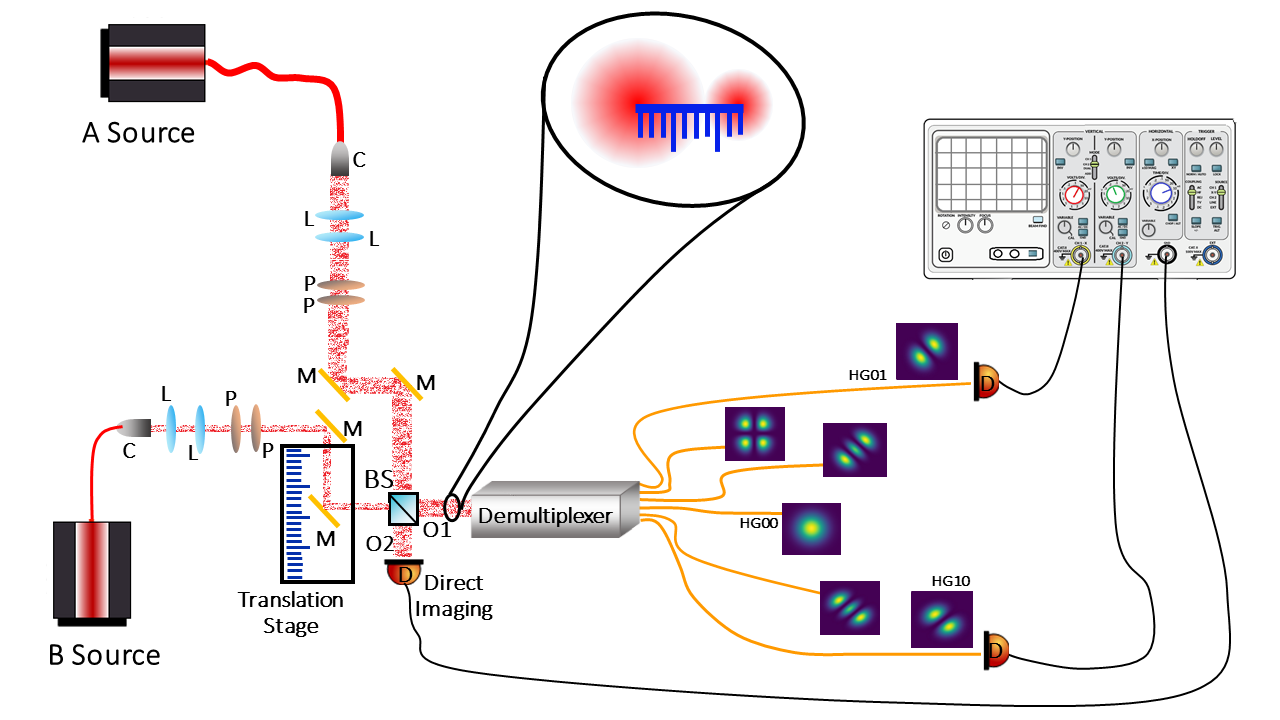}
    \caption{Two telecom fiber lasers (A and B sources) that exit from collimators (C) are mode-matched using simple lenses (L) systems.
The intensities are tuned by changing the relative orientation of a pair of polarizers (P).
The beams are combined on the input ports of a non polarizing beam splitter (BS) whose O1 output port is coupled with the demultiplexer input free space port.
A pair of steering (M) mirrors for each beam are used to optimize the coupling with the demultiplexer. The second mirror of beam B is mounted on a translation stage to move the beam, within transverse plane, with micrometric resolution whereas the A beam stays centered throughout the measurement. 
The demultiplexer, PROTEUS-C from Cailabs, allows to perform intensity measurements on six Hermite-Gaussian mode.
The $\text{HG}_{01}$ and $\text{HG}_{10}$ outputs of the demultiplexer are coupled with photodetectors (D) whose signals $F_1$ and $F_2$ are recorded using an oscilloscope.
The intensity of the second output (O2) of BS is recorded using the same detector/oscilloscope used to record intensities of  $\text{HG}_{01}$ and $\text{HG}_{10}$ modes.}
    \label{setup}
\end{figure}


\section{Experimental setup}\label{Sec:exp}

We perform imaging of two point-like sources, at variable distance ($d$) and intensity ratio ($\epsilon = I_B/I_A$), using spatial demultiplexing in the HG modes.
We use a demultiplexer, PROTEUS-C from Cailabs, based on Multi-Plane Light Conversion. 
We combine two independent (to avoid interference) telecom fiber lasers (1.55 $\mu m$) of different power on a non-polarizing beam splitter (NPBS) to mimic two point-like sources. 
We denote as A the more intense beam, whereas B is the weaker one. 
As shown in figure \ref{setup} each fiber laser is coupled with a collimator and mode-matched using two lenses to get a beam waist $w_0 \simeq 300 \mu m $ as required by the demultiplexer specifications.  
Two polarizers are used to change the beams power in a controlled way.
The beams  are combined on the NPBS through a pair of steering mirrors that are also used to couple each beam in the input (free-space) port of the HG  demultiplexer.

The second mirror of the lower power B beam (adjustable between 0 and 11 $\mu W$) is mounted on a translation stage to control its location, within the transverse plane, with micrometer resolution.
The demultiplexer allows us to perform intensity measurements on six HG modes. It accepts radiation from the free-space input port, and decomposes it in the lowest-order modes ($\text{HG}_{00}$, $\text{HG}_{01}$, $\text{HG}_{10}$, $\text{HG}_{11}$, $\text{HG}_{02}$, $\text{HG}_{20}$). The modes are coupled with six single-mode fibers following conversion into the $\text{HG}_{00}$ mode.
Finally, the modes $\text{HG}_{01}$ and $\text{HG}_{10}$ impinge on two nearly identical commercial InGaAs photodiodes, whose signals are recorded on an oscilloscope.
A removable power meter is used to measure the power of the beams for calibration before impinging on  the demultiplexer.


\begin{figure}
    \centering
    \includegraphics[scale=.35]{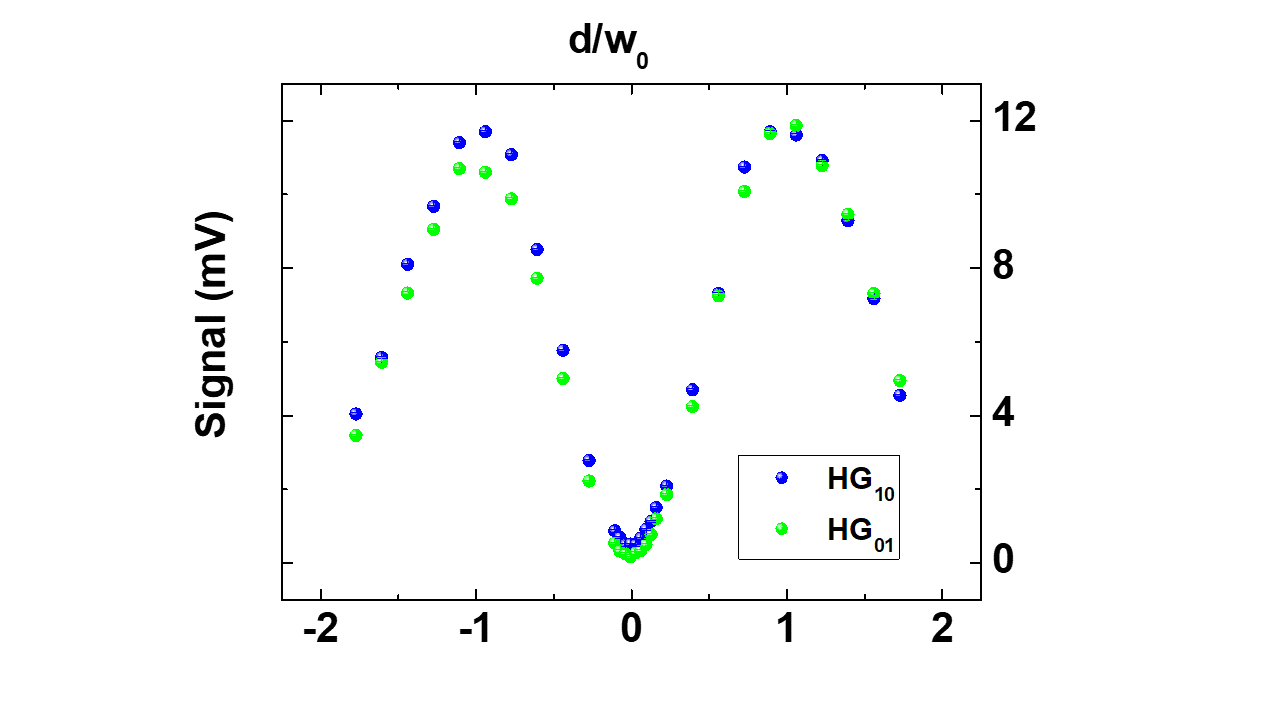}
    \caption{Measured signals $F_1$ and $F_2$ (corrresponding to $\text{HG}_{10}$ and $\text{HG}_{01}$ modes) at different values of separation $d/w_0$ between simulated A and B sources.}
    \label{figura2}
\end{figure}


\subsection{Alignment}

When the translation stage is centered, the beams separation is zero (i.e., the beams overlap completely) and the overall source (A+B beam) has a circular symmetry. 
In this case the powers leaked in the $\text{HG}_{01}$ and $\text{HG}_{10}$ modes are expected to reach the minimum value, determined only by cross-talk.
We use this to calibrate the setup.
For the calibration, we first align the device by maximising the signal in the $\text{HG}_{00}$ mode when the beams are completely superimposed.
Then we minimize the power in the outputs of $\text{HG}_{10}$ and $\text{HG}_{01}$ ports with tiny movements on the coupling mirrors, being careful not to change the power in $\text{HG}_{00}$ mode.
Data are collected by keeping the A beam centered and translating the B beam.
However, in a practical scenario one can only hope to align the demultiplexer with the centroid, since the positions of the sources are unknown.
This means that the correct procedure to calibrate the device should be to maximize the $\text{HG}_{00}$ signal as we did, but then, to translate the beams in opposite directions.
Anyway, in our case, being the translated beam much weaker than the strong centered beam, our procedure, consisting in translating only one beam (B beam), does not affect the calibration.


\begin{figure}
    \centering
    \includegraphics[scale=.35]{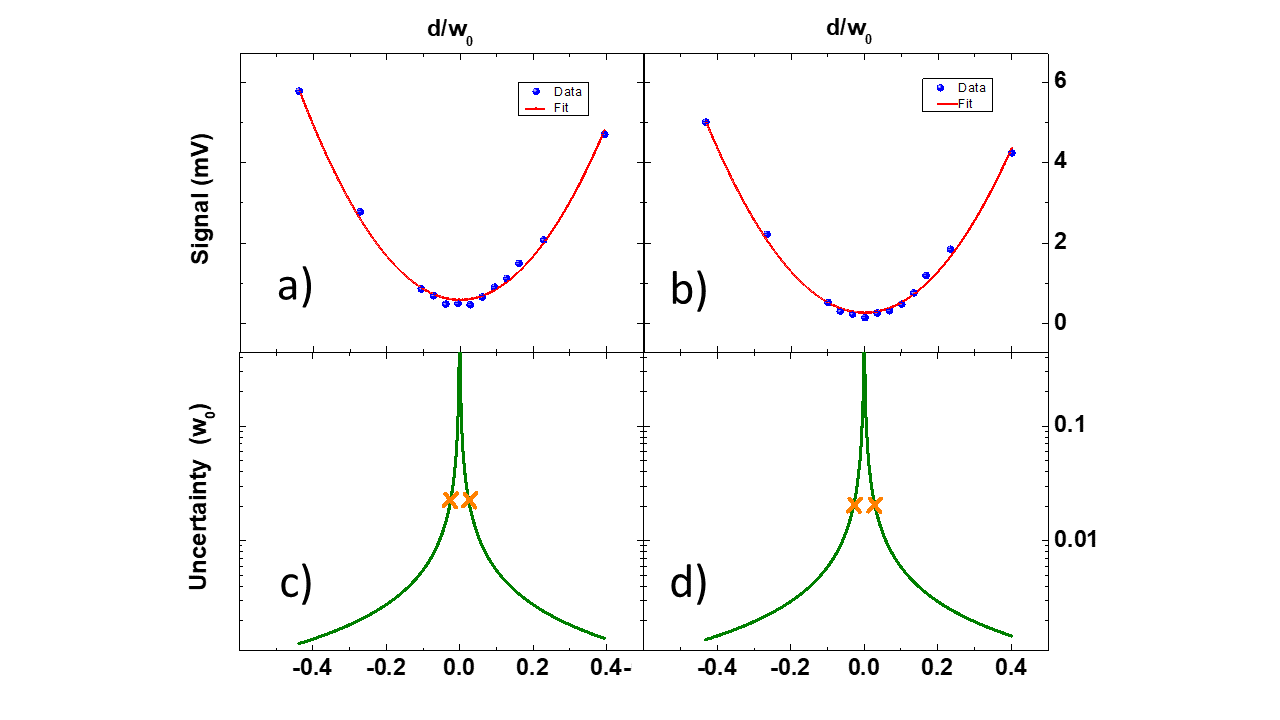}
    \caption{Measured signals $F_1$ and $F_2$ (upper plots, blue points) at different values of separation $d/w_0$ between simulated sources in quadratic range ($d/w_0<1$) and quadratic fits (red curves).
    Using these fits as calibration curves, it is possible to estimate the beam separation by simply measuring $\overline{F_1}$ or $\overline{F_2}$ and by finding the corresponding beam separation $d\cos{\theta}=F_1^{-1}( \overline{F_1} )$ and $d\sin{\theta}=F_2^{-1}( \overline{F_2} )$.
    In the lower graphs, the green curves represent the uncertainties $\delta d /w_0$ associated to the estimation of the beam separation $d/w_0$ using the upper graphs as calibration curves and uncertainty determination procedure as explained in the text. The orange crosses represent the resolving power (when $\delta d \simeq d$).
The resolving power are nearly the same ($r \simeq 0.023$) both for $\text{HG}_{10}$ and $\text{HG}_{01}$ mode,  confirming the symmetry of the demultiplexer, and the reliability of the alignment.}
    \label{figura3}
\end{figure}


\subsection{Separation measurement}

Our first goal is to estimate the trasverse separation $d$ between the two beams, each having beam waist $w_0 \simeq 300 \mu m$.
To calibrate the setup, we measured the photodiode signal $F_1$ ($F_2$) on the oscilloscope corresponding to intensity of $\text{HG}_{01}$ ($\text{HG}_{10}$) for varying values of the separation $d$.
We allow $d$ to assume both positive and negative values, where positive (negative) values correspond the weak beam being to the left (right) of the strong beam.
Using these signals as calibration curves, it is possible to estimate the beam separation by simply measuring $F_1$, $F_2$ and then finding the corresponding beam separation $d$.

Figure \ref{figura2} shows the measured signals $F_1$ and $F_2$ as functions of $d$.
The strong (A) beam has a power of $150 \mu W$, the B beam of $11 \mu W$, the integration time is $100 ms$ with an oscilloscope sampling rate of $2.5$ MegaSample per second.
In these experimental conditions, errors due to integration time are negligible, and statistical errors due to repeatability of the measurement dominate.
In detail, to estimate the error bars, we perform repeated measurements of the voltage on the oscilloscope. For each measurement, we change the position of the translation stage, then return to original position and finally acquire again the voltage on oscilloscope. 
We repeated the procedure eleven times to estimate the standard deviation $\delta F_i$ ($i=1,2$), which we use to quantify the uncertainty in the estimation of $d$.
Note that this procedure to estimate the uncertainty takes into account all possible error sources, including oscilloscope discretisation error, detector voltage noise and offset fluctuations, laser fluctuations and repeatability of the translation stage 
(changes in the position of the translation stage when its position is approached from different directions).

To establish the ultimate limit of this measurement procedure, we focus on the most interesting regime of small values of the separation $d$, i.e.,~for  $d \ll w_0$.
We used the data $F_1$ and $F_2$, shown in figures \ref{figura3}a) and \ref{figura3}b) to perform two parabolic fits ($F_1(d)$ and $F_2(d)$), as suggested from the theory in the sub-Rayleigh regime.
We obtain reassuring values for the reduced $\chi^2$ ($\chi^2(F_1)=0.014$ and $\chi^2(F_2)=0.015$). 
The small values of $\chi^2$ confirm that $F_1(d)$ and $F_2(d)$ are well approximated by quadratic functions, \begin{align}
F_1(d) 
& \propto d^2 + \text{const} \, , \\
F_2(d) 
& \propto d^2 + \text{const} \, .
\end{align}

We quantify the uncertainty $\delta d$, on the estimated separation $d$ in units of the beam waist $w_0$, by means of the calibration curves $F_1(d)$ and $F_2(d)$ through the following procedure:
\begin{enumerate}
\item Inversion of the $F_i(d)$ (for $i=1,2$) fitted curves shown in figure \ref{figura3}a) and \ref{figura3}b), $F_i(d) \to d(F_i)$;

\item Calculation of the derivative $\partial d(F_i) / \partial F_i$ using the theoretical curve;

\item Propagation of the error on the measured signal $\delta F_i$,
which finally yields
$\delta d \simeq \delta F_i \left| \partial d(F_i) /\partial F_i \right|$.
\end{enumerate}

Figure \ref{figura3}c) and \ref{figura3}d) show the calculated uncertainty as function of the beams distance, which, as expected, diverge at zero separation. 
When the uncertainty $\delta d$ is equal to the separation ($\delta d = d$) we reach the resolving power {$r=d_m / w_0$ of our setup, where $d_m$ is the minimum detectable separation.
Below this value it is virtually impossible to resolve the two sources.
This limiting point is indicated by the two orange crosses. 
The obtained resolving power is about $2.4\%$ of the beam waist.
As expected, uncertainty decreases away from the sub-Rayleigh regime, where we obtain an uncertainty on the separation of about $0.1\%$. 
Moreover, the same values of uncertainty and resolving power are obtained for $\text{HG}_{10}$ and $\text{HG}_{01}$ in either direction (A source to the left or to the right of the B source) confirming the reliability of the alignment. 
For comparison, Boucher et al.~\cite{Boucher2020} reported a minimum detectable separation of about $5 \%$
using a similar Multi-Plane Light Conversion system \cite{Morizur2010, Labroille2014}, but with two sources of equal intensity and about two orders of magnitude more intense than in our setup.


\begin{figure}
    \centering
    \includegraphics[scale=.35]{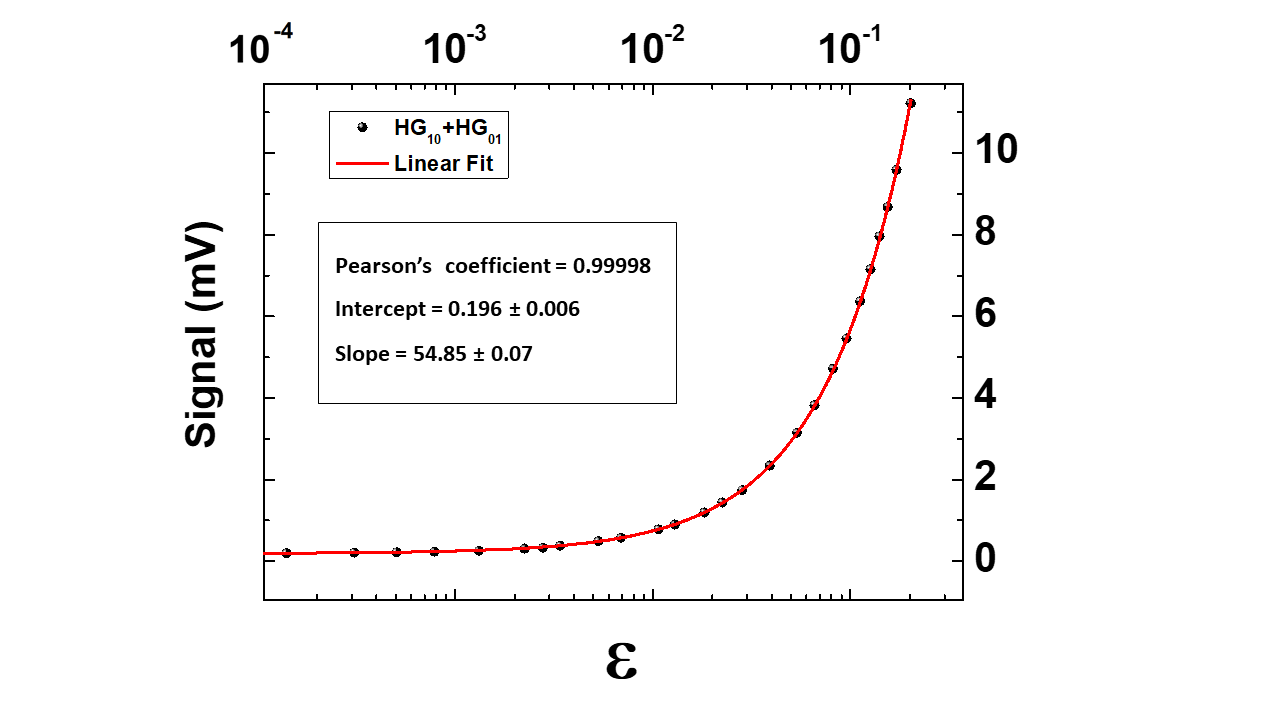}
    \caption{Intensity Ratio ($\epsilon$) measurement using HG SPADE. Sum of the signal $F_1$ and $F_2$ at different values of $\epsilon$ (black points) and corresponding linear fit (red curve). The strong A source  ($33.6 \mu W$) is aligned with the demultiplexer optics and the weak B sources, fixed at distance $d = \frac{2}{3} w_0$, is progressively attenuated to change $\epsilon$.}
    \label{figura4}
\end{figure}


\subsection{Measure of the Intensity Ratio}

Our second goal is to estimate the intensities ratio $\epsilon$ between the two sources.
The strong A source, now $33.6 \mu W$, is aligned with the demultiplexer optics and the weak B sources is fixed at distance $d = \frac{2}{3} w_0$.
To obtain the calibration curve, we record the signal $V$ given by the sum of $\text{HG}_{10}$ and $\text{HG}_{01}$ intensities ($V = F_1 + F_2$) for different intensity values of the weak source. The measurement results are shown in figure \ref{figura4}.

As for the case of distance metrology, repeatability errors are the dominant sources of uncertainty in this setup.
To estimate the repeatability error, we proceed as in the previous case: we acquire the voltage on the oscilloscope, then change the $\epsilon$ value, then return to original value and finally acquire again the voltage on oscilloscope. %
We repeat this measurement ten times to obtain an estimate of the standard deviation $\delta V$, which is equal to $1.5 \mu V$ at the detection limit.

The signal is proportional to the intensity of the B source plus an offset due to cross talks $x_{10}^{00}$ and $x_{01}^{00}$. The strong beam, even if perfectly centered, leaks power in $\text{HG}_{10}$ and $\text{HG}_{01}$ modes, setting the detection limit $\epsilon_\ell$.
More exactly, $\epsilon_\ell$ is determined by the total fluctuations.
However, being the fluctuations of B negligible compared to the fluctuation of A in regime of small $\epsilon$, the detection limit is determined by the fluctuation of the A source.
We performed a linear fit (red line in figure \ref{figura4}) obtaining a Pearson coefficient larger than 0.9999, which confirms the linearity of the system.
As for the case of distance metrology, we use the fitted law $V(\epsilon)$ as a calibration curve. 
This yields our estimate of the error in the estimation of $\epsilon$, $\delta \epsilon \simeq \delta V \left| \partial \epsilon(V) /\partial V \right|$.

In the limit of $\epsilon \to 0$ we obtain $\delta \epsilon \rightarrow \epsilon_\ell \simeq 2.7 \cdot 10^{-5}$. 
This means that the setup would allow us to 
determine the presence of a secondary source that is $10^5$ times weaker than the primary source.
Note that the obtained value for $\epsilon_\ell$ depends on the power fluctuations of the strong A source, which in turn depend on the power of A itself. 
For a shot-noise limited source, the fluctuations go as the square root of the power, therefore, the detection limit $\epsilon_\ell$ is expected to decrease for increasing power of the primary source. 

To obtain a quantitative comparison with DI, we performed a measurement for determining the detection limit $\epsilon_\ell^{\text{DI}}$ for DI using the same source powers, distance, waists, oscilloscope, integration time and sampling rate. However, we consider only one pixel, located in correspondence of point of maximum intensity of the secondary source. 

We coupled the output O2 of the beam splitter in figure \ref{setup} with the same detector used for SPADE measure to record the total intensity (A+B beams).
We recorded the detector signal on the oscilloscope for different value of $\epsilon$ keeping the power of A fixed (figure \ref{figura5}). 
Also in this case the error bar are determined by the repeatability error and we performed a linear fit of the experimental points producing the calibration curve (red curve in figure \ref{figura5}).
Using the same procedure used for the data in figure \ref{figura4} to estimate the detection limit (inversion, derivative, multiplication for the signal error), we obtain a detection limit for $\epsilon$ in direct imaging case $\epsilon_\ell^{\text{DI}} \simeq 5.7 \cdot 10^{-4}$ nearly 21 times worse than SPADE.

For DI, when the beam waist is larger than the pixel size, as is the case in our setup, and the distance $d$ between the two beams is smaller than the beam waist, roughly the same amount light intensity from A and B impinge on the detector, determining a cross talk (intercept in the inset of figure \ref{figura5}) much larger than the cross talk of HG SPADE. This explain the worse performance of DI compared to SPADE.


\begin{figure}
    \centering
    \includegraphics[scale=.35]{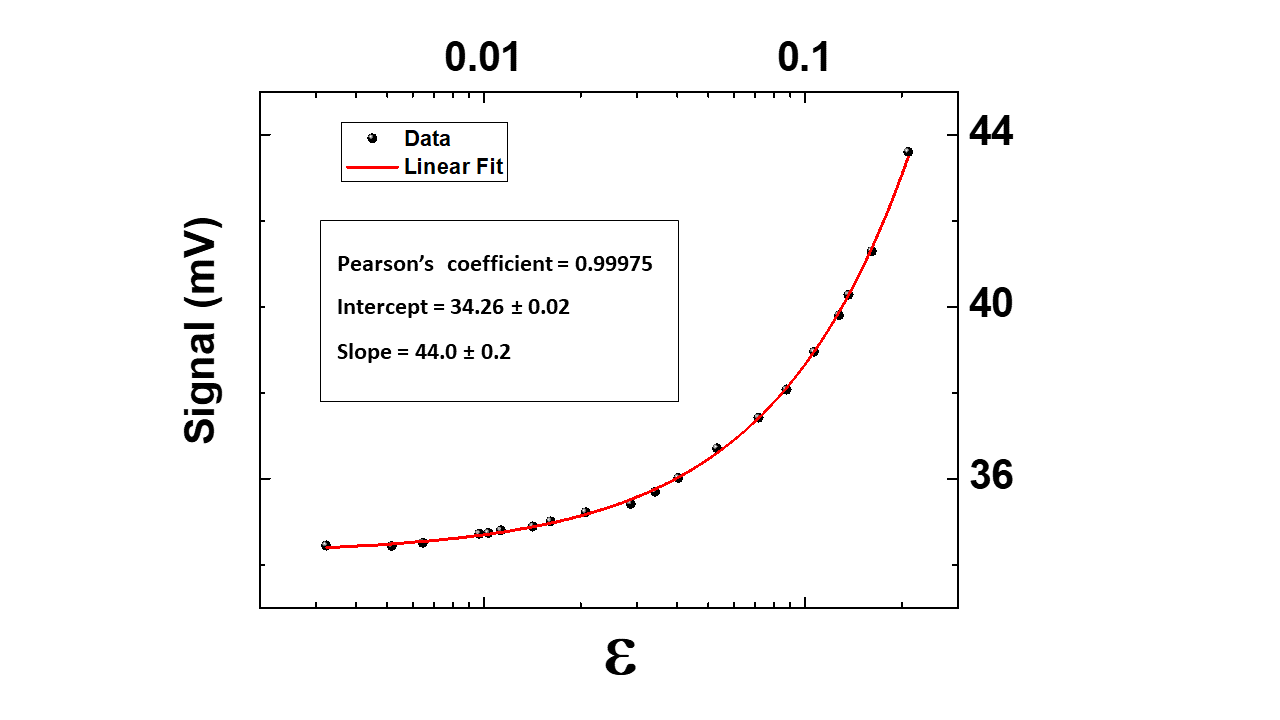}
    \caption{Intensity Ratio ($\epsilon$) measurement using DI. Signal measured on O2 output port of the beam splitter at different values of $\epsilon$ (black points) and corresponding linear fit (red curve). The strong A source  ($33.6 \mu W$) is aligned with the demultiplexer optics and the weak B sources, fixed at distance $d = \frac{2}{3} w_0$, is progressively attenuated to change $\epsilon$.}
    \label{figura5}
\end{figure}


\section{Conclusions}

Spatial mode demultiplexing (SPADE) allows us to achieve the ultimate quantum limit in the estimation of the transverse separation between two weak incoherent sources, which emits no more than one photon per detection window~\cite{Tsang2016}. 
This performs much better than direct imaging (DI), as the latter is not able to capture the information encoded in the phase of the field.
More generally, SPADE is optimal for estimate the second moments of the intensity distribution of a source of arbitrary shape in the weak signal limit~\cite{Modern}. In fact, in the sub-wavelength regime it allows to estimate the second moments with a single intensity measurement.

In this work we have analysed the use of SPADE in the regime of bright signals. First we have shown that the signal-to-noise ratio is expected to be much larger for SPADE than for DI, though SPADE is limited by cross talk between different spatial channels.
For the case of two point-like sources of different brightness, our theory is complemented by an experimental demonstration of the use of SPADE for the estimation of the sources transverse separation and of their relative intensity. 
This paves the way to potential applications of SPADE even in the regime of bright signals, for image recognition~\cite{GraceGuha,Shapiro} and astronomical observations~\cite{PRL2021,astro22} with sub-wavelength precision.

\vspace{0.5cm}
\noindent
\textbf{Funding.}
We gratefully acknowledge support from the Italian Space Agency (ASI) through the Nonlinear Interferometry at Heisenberg Limit (NIHL) project (CUP F89J21027890005);
and from the 
European Union – Next Generation EU: NRRP Initiative, Mission 4, Component 2, Investment 1.3 – Partnerships extended to universities, research centres, companies and research D.D. MUR n. 341 dated 15.03.2022 (PE0000023 – ``National Quantum Science and Technology Institute").\\
\textbf{Acknowledgments.}
We thank Cailabs, 38 boulevard Albert 1er, 35200 Rennes, France.\\
\textbf{Disclosures.}
The authors declare no conflicts of interest.\\
\textbf{Data Availability Statement.}
Data available from the authors on request.\\
\textbf{Authors contribution.}
L.S.A. designed the experimental setup; C.L developed the model for the error analysis; L.S.A., D.K.P, M.S.d.C. and D.D. realized the apparatus, L.S.A performed the data acquisition and analysis; C.L assisted with the experimental design and supervised the experiment, all authors discussed the results and contributed to the final manuscript.

\end{document}